\documentclass[%
reprint,
 amsmath,amssymb,
 aps,
]{revtex4-2}

\usepackage{subfigure}
\usepackage{float}
\usepackage{slashed}
\usepackage{xcolor}
\usepackage{amsmath}
\usepackage{subcaption}
\usepackage[english]{babel}
\usepackage{tikz}
\usepackage{adjustbox}
\usepackage{rotating}
\usepackage{rotfloat}
\usepackage{verbatim}
\usepackage[utf8]{inputenc}
\usepackage{graphicx}
\usepackage{dcolumn}
\usepackage{bm}
\usepackage[super]{nth}
\usepackage{lipsum}

\usepackage{hyperref}
\usepackage{xcolor}
\hypersetup{
  colorlinks   = true, 
  urlcolor     = red, 
  linkcolor    = blue, 
  citecolor   = blue 
}

\usepackage{threeparttable}
\begin{document}

\preprint{APS/123-QED}

\title{Search for the production of dark Higgs in the framework of Mono-Z$^{\prime}$ portal at the FCC-ee simulated electron-positron collisions at $\sqrt{s} = 240$ GeV}

\author{S. Elgammal}
 \altaffiliation[sherif.elgammal@bue.edu.eg]{}
\affiliation{%
Centre for Theoretical Physics, The British University in Egypt, P.O. Box 43, El Sherouk City, Cairo 11837, Egypt.
}%

\author{N. De Filippis}

\affiliation{%
Istituto Nazionale di Fisica Nucleare, Sezione di Bari: Bari, Apulia, IT.
}%

\affiliation{%
Politecnico di Bari: Bari, Apulia, IT.
}%


\date{\today}

\begin{abstract}
{We conducted a Monte Carlo search for dark Higgs particles ($h_{D}$), which is based on a model derived from the Mono-Z$^{\prime}$ portal. This model suggests the existence of $h_{D}$ and a new neutral gauge boson, denoted as Z$^{\prime}$. 
Our study focused on the event topology featuring two oppositely charged muons and missing energy, generated from simulations of $e^+ e^-$ collisions at the Future Circular Collider (FCC-ee). This collider is set to operate with 240 GeV a
center-of-mass energy ($\sqrt{s}$), and will achieve an integrated luminosity of 10.8 ab$^{-1}$. In our analysis, we estimated upper limits on the mass of the dark Higgs at a 95\% confidence level, assuming that no $h_{D}$ particles were detected.}

\end{abstract}

\maketitle

\section{Introduction}
\label{sec:intro}
The observation of the Standard Model (SM) Higgs boson at the Large Hadron Collider (LHC) in 2012 filled a significant gap in the SM \cite{HiggsATLAS,HiggsCMS}. 
Researchers are still analyzing its properties, but to achieve precise results, future $e^+ e^-$ colliders are needed to generate vast amounts of Higgs bosons. Accurate calculations could provide valuable insights into new BSM physics.

A lot of theories propose that the Higgs boson could serve as a gateway to a new understanding of particle physics, especially through the decays of exotic particles into new light particles \cite{DH-at-colliders, Higgs1, Higgs2, Higgs3, R1, atlasmonoZprime}. Therefore, it is crucial to search for these exotic decays alongside conducting precision studies at future Higgs factories \cite{Higgs4}. Furthermore, non-minimal dark sectors, which contain additional hidden particles or exhibit cascade decays, impose limits that are highly dependent on various assumptions regarding the kinetic mixing parameter, branching ratios, and decay topologies \cite{DS-at-colliders}.

While decays into particles that decay promptly have been explored at the LHC and proposed future facilities, new physics may also emerge in other decay modes \cite{monoHiggsAtlas1, monoHiggsAtlas2}. 

The ATLAS \cite{HiggsAtlas24,HiggsAtlas25,HiggsAtlas26,HiggsAtlas27,HiggsAtlas28,HiggsAtlas29} and CMS \cite{HiggsCMS30,HiggsCMS31,HiggsCMS32,HiggsCMS33,HiggsCMS34,HiggsCMS35,HiggsCMS36} collaborations have conducted direct studies for the decay of the Higgs boson to invisible particles. These investigations were based on the collected data during Run 1 (2011–2012) and Run 2 (2015–2018). 
These analyses focused on Higgs boson production through the fusion of gluon-gluon,  vector boson (VBF), and Higgs coupling to vector boson as VH or top quark pair as ttH. 

The CMS experiment set the most stringent limit of about 0.18 on the branching fraction Br($H \rightarrow \text{inv}$) at 95\% confidence level (CL), using VBF data from LHC Runs 1 and 2, while the expected value was 0.10 \cite{HiggsCMS36} with the Higgs mass fixed at 125 GeV.
Similar searches have been performed in previous experiments such as LEP; no evidence was found for the invisible Higgs alongside Z bosons decaying to hadrons, electrons, or muons. Based on these analyses, the achieved limits on invisible Higgs mass range from 50 - 112.1 GeV from hadronic decays and 50 - 91.3 GeV from leptonic decays \cite{LEP-Higgs}.

One of the methods for searching for physics beyond the SM (BSM) at future $e^+ e^-$ colliders is analyzing the distribution of dilepton mass, looking for significant changes. These changes may indicate new peaks from neutral gauge bosons like \( Z' \) \cite{heaveyZ}, distortions from Contact Interactions \cite{ContactInteraction}, or features from models such as ADD and Randall-Sundrum \cite{ADD, extradim}. 
Recently published results by CMS have excluded Z$^{\prime}$ at a 95\% CL for masses between 0.6 and 5.15 TeV \cite{ZprimeandCI}; ATLAS has ruled out masses from 0.6 to 5.1 TeV \cite{zprimeATLAS}.

The future $e^+ e^-$ colliders, such as the proposed FCC-ee, will be vital for studying BSM. The FCC-ee is designed to start operations at $\sqrt{s} = 240$ GeV, with plans for an upgrade to 365 GeV \cite{FCC-ee1, FCC-ee2, FCC-ee3}. The initial phase of the FCC-ee is designed as a high-luminosity $e^+ e^-$ collider, focused on producing electroweak interactions, top quark events, and Higgs boson processes. This setup allows for adjustable energy levels, helping to effectively reduce the QCD background \cite{FCC-ee4}.

This analysis investigates a light dark Higgs ($h_{D}$) produced at FCC-ee alongside light Z$^{\prime}$ 
boson with mass $M_{Z^{\prime}} < 80$ GeV, using DH simplified model in a Mono-Z$^{\prime}$ portal \cite{R1}. We focus on simulated $e^+ e^-$ collisions at $\sqrt{s} = 240$ GeV, examining dimuon decay of Z$^{\prime}$ and missing transverse energy ($E^{miss}$) from dark Higgs decaying to invisible dark matter (DM).

This paper is organized as follows: 
In Section \ref{section:constraints}, we present plots that showcase the exclusion limits pertinent to our study, drawing on insights from previous experiments and various FCC-ee investigations. Section \ref{section:model} outlines the theoretical framework of the Mono-Z$^{\prime}$ portal model. Meanwhile, Section \ref{section:MCandDat} delves into the simulation methods we utilized to generate both signal and background samples. In Section \ref{section:AnSelection}, we detail the selection criteria and analysis strategy employed, while Sections \ref{section:Results} and \ref{section:Summary} wrap up by summarizing our findings and the overall analysis.

\section{Current limitations on the parameter space}
\label{section:constraints}
\begin{figure}
\centering
\subfigure[]{
  \includegraphics[width=63mm]{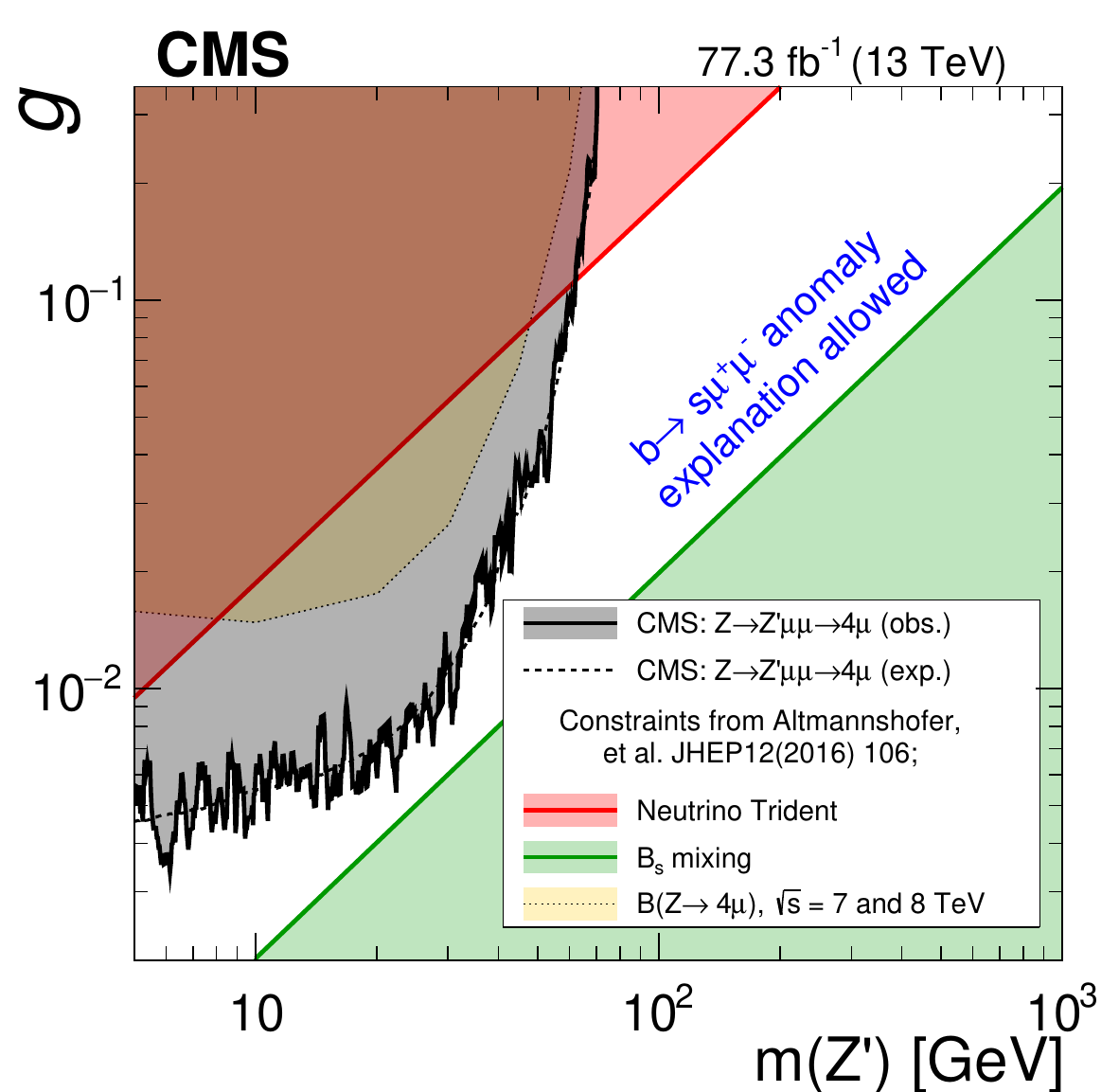}
  \label{cms-cons}
}
\subfigure[]{
  \includegraphics[width=65mm]{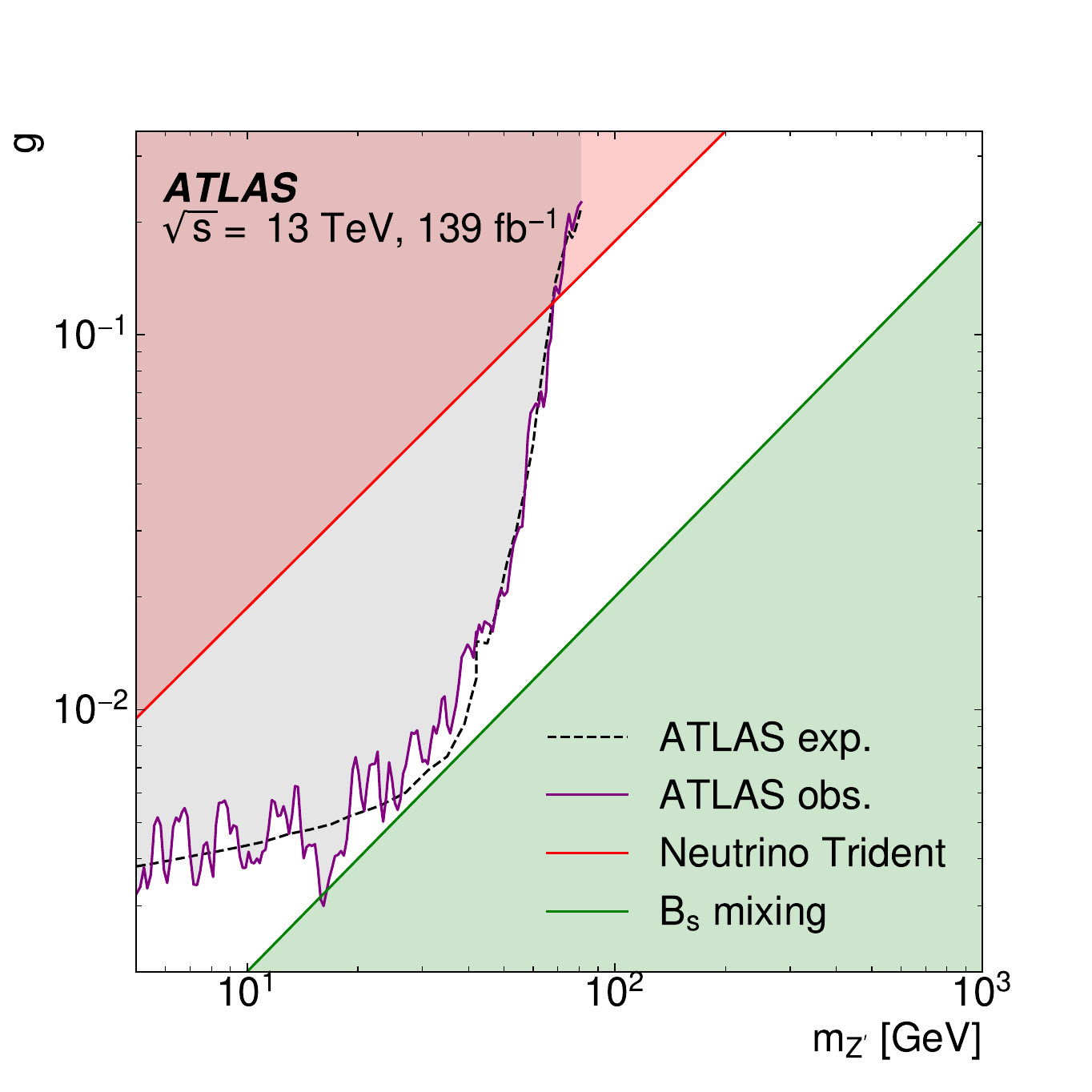}
  \label{atlas-cons}
}
\caption{The coupling parameter $\text{g}_{l}$ limits from the CMS search \cite{R130} as a function of the $Z'$ mass compared to the ATLAS experimental results \cite{R131}.
}
\label{constraints:CMSandATLAS}
\end{figure}

Searches for a \( Z' \) gauge boson, derived from an \( L_\mu - L_\tau \) U(1) local gauge symmetry, have been carried out using data from proton-proton collisions at a center-of-mass energy of \( \sqrt{s} = 13 \) TeV, corresponding to the LHC Run 2. These analyses were conducted by the CMS \cite{R130} and ATLAS \cite{R131} collaborations. The events of interest involved four muons with an invariant mass close to that of the Standard Model Z boson. To enhance the search sensitivity, they focused on the decay channel \( Z \rightarrow Z' \mu\mu \rightarrow 4\mu \). As a result, they were able to place constraints on the coupling strength of the \( Z' \) boson to muons (\( \text{g}_{l} \)), ruling out at 95\% CL values exceeding 0.004 to 0.3, as illustrated in Figure \ref{constraints:CMSandATLAS}.

The direct search for new physics through \( Z' \) decays into dileptons will yield a distinct signal at the FCC-ee. A detailed study on detecting a leptophilic \( Z' \) at FCC-ee has been presented in \cite{Zprime-couplin-to-ll}, which predicts the production of \( Z' \) along with initial-state radiation ($Z' \gamma$). This research indicates that within the mass range of 10 to 365 GeV, the FCC-ee is expected to enhance the exclusion limits on the coupling $\text{g}_{l}$ by one to two orders of magnitude. 

\begin{figure*}
\centering
\resizebox*{13.cm}{!}{\includegraphics{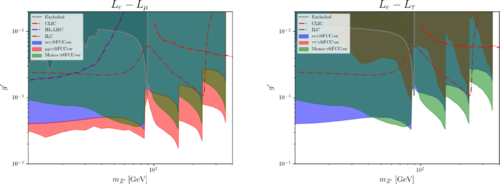}}
\caption{Projected sensitivities of the FCC-ee to the $L_e$ - $L_\mu$ ($L_e$ - $L_\tau$) models. The red, blue, and green bands correspond to the $\mu\mu\gamma$ ($\tau\tau\gamma$), $ee\gamma$, and mono-$\gamma$ search channels, while dot-dashed
lines show the projected reach of the HL-LHC (purple), respectively, and show the strongest bounds set on this model by each search channel \cite{Zprime-couplin-to-ll}.}
\label{constraints:hllhc}
\end{figure*}

As illustrated in Figure \ref{constraints:hllhc}, the parameter space that would ultimately be excluded falls around $\text{g}_{l} \sim 3(4) \times 10^{-4}$ for the $L_e - L_\mu$ and $L_e - L_\tau$ scenarios. In contrast, the HL-LHC begins to lose sensitivity for $\text{g}_{l} < 0.002$. 

In a novel approach complementary to this study, we aim to investigate the production of \( Z' \) alongside an invisible Higgs particle within the framework of Mono-Z$^{\prime}$.

Finally, older $e+~e-$ experiment as the LEP experiment \cite{lep} provides a constraint of \( \text{g}_{l} \leq 0.044 \times M_{Z^{\prime}}/(\text{200 GeV}) \) for $M_{Z'} > 209$ GeV and \( \text{g}_{l} \leq 0.044 \) for \( M_{Z^{\prime}} < 209~\text{GeV} \) \cite{R37}.

\section{The theoretical model of Dark Higgs in the framework of the Mono-\( Z' \)}
\label{section:model}
The Mono-\( Z' \) framework \cite{R1,R222,R333} offers a streamlined depiction of interactions within the dark sector, where a new neutral gauge boson, \( Z' \), serves as a mediator between the SM and the dark sector. This mediator can naturally emerge in SM extensions that include an additional $U(1)^{\prime}$ gauge symmetry. In scenarios where this new gauge symmetry is spontaneously broken, a related dark Higgs boson may appear, playing a crucial role in generating the \( Z' \) mass. 
The dark Higgs can decay dominantly to the invisible $\chi$ states through the $\lambda_D$ coupling, which we can take to be O(1) \cite{R1}. 
Therefore, investigating dark Higgs production alongside missing energy provides an essential avenue for exploring the dark sector and its symmetry-breaking mechanisms.

The model under consideration is discussed in \cite{R1}; it is considered one of the top recommendations from the CMS collaboration for searching for dark Higgs at the LHC \cite{DMrecommendationsByCMS}. 
In this model, DM can be generated via three scenarios: Light vector (LV), EFT, and Dark Higgs (DH). 
We studied the production mechanism of the DH scenario at the FCC-ee collider, as well as that of the Z$^{\prime}$ boson.

\begin{figure}
\centering
\resizebox*{8.cm}{!}{\includegraphics{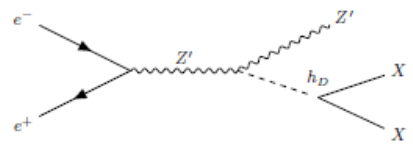}}
\caption{The DH scenario Feynman diagram as modified from \cite{R1}.}
\label{figure:fig1}
\end{figure}

The Z$^{\prime}$ interacts with SM leptons and the dark Higgs particle. It generates a light $h_{D}$, which decays to two particles ($\chi$ and $\bar{\chi}$) of dark matter. 
The DH scenario Feynman diagram is illustrated in Figure \ref{figure:fig1}.

Interaction part, in the Lagrangian, between the dark Higgs and Z$^{\prime}$ is given by \cite{R1} 
\begin{equation}
    \text{g}_{D} M_{Z^{\prime}}h_{D} Z'^{\mu} Z'_{\mu}, \nonumber
\end{equation}
where $\text{g}_{D}$ denotes the coupling of the DH to Z$^{\prime}$.

While the interaction part in which the SM particles can couple to Z$^{\prime}$ is presented in \cite{R1}
\begin{equation}
    \sum_{l} \texttt{g}_{l} \bar{\psi_{l}}\gamma_{\mu}\psi_{l} Z'^{\mu}. \nonumber
\end{equation}

\begin{table}
\centering
\begin{tabular}{|l | l|}
\hline
\textbf{Scenario} & \textbf{Mass assumptions} \\
\hline
Light $M_{h_D}$ & $M_{h_D} = \begin{cases}
                   M_{Z'}\,, & M_{Z'} < 125\,\mathrm{GeV}\\
                  125~\mathrm{GeV},            & M_{Z'} > 125\,\mathrm{GeV}.
\end{cases}$
                  \\
\hline
\end{tabular}
\caption{Mass assumptions of the light $M_{h_D}$ particles produced in the DH model as outlined in Ref.~\cite{R1}.
}
\label{tab:mass_assumptions}
\end{table}

In this scenario, the masses of the dark Higgs ($M_{h_{D}}$) and the Z$^{\prime}$ are treated as independent variables within the model, although they are both determined by the same scale, a vev ($\nu_D$). The relationship between their masses is given by $M_{h_{D}} / M_{Z^{\prime}} = \sqrt{2}\lambda_D/\text{g}_{D}$ \cite{R1}. It is important to note that the $Z' + h_{D}$ signal, as illustrated in Figure \ref{figure:fig1}, shows a preference for larger values of $\text{g}_{D}$. This suggests that, for perturbative couplings, the dark Higgs cannot substantially exceed the mass of the $Z'$ \cite{R1}. 
To effectively capture the range of effects produced by varying particle masses, we focus on two benchmark scenarios, which are detailed in Table \ref{tab:mass_assumptions}.

In the current analysis, we consider the mass assumption at which \(M_{h_{D}} = M_{Z^{\prime}}\) for \(M_{Z^{\prime}} < 125\) GeV to prevent overlapping with the SM Higgs.

In this scenario, the allowed decay particles are the off-shell ${Z}'\rightarrow h_{D} Z'$, $h_{D}\rightarrow\chi\chi$, and on-shell ${Z}'\rightarrow\mu^{+}\mu^{-}$. 
Meanwhile, decay widths of $h_{D}$ and ${Z}'$ were computed using their masses and coupling constants. 
In this scenario, the cross section calculations depend on the masses of the DH and mediator, and not on the DM mass \cite{R1}.

Because of earlier limits from ATLAS, CMS, and LEP-2 experiments on the coupling \(\text{g}_{l}\), the value of \(\text{g}_{l}\) must be less than 0.003 for \(M_{Z^{\prime}}\) within the range of 10 to 80 GeV \cite{R37}. This range is of particular interest for our analysis. 
This constraint also falls within the sensitivity range established in \cite{Zprime-couplin-to-ll} for the Mono-$\gamma$ analysis at FCC-ee. 

In this context, the value of \(\texttt{g}_{D}\) is fixed at 1.0, as advised by the ATLAS collaboration for the search involving dark Higgs at the ATLAS detector \cite{atlas-Zp}. Consequently, the parameters that remain free in this scenario include the masses (\(M_{Z^{\prime}}\), \(M_{h_{D}}\)) along with the coupling constant \(\texttt{g}_{l}\).

\section{The Monte Carlo simulation of DH signal samples and SM backgrounds}
\label{section:MCandDat}
The signal samples for the DH scenario and related SM background samples have been privately generated using Monte Carlo (MC) simulations. 
Both were produced via the WHIZARD \cite{whizard} event generator with version 3.1.1. To account for the initial state radiation (ISR) effect, this was integrated using Pythia 6.24, which provides hadronization \cite{R34} and handles the parton shower modeling. 
The DELPHES package \cite{delphes} was employed for the IDEA detector fast simulation \cite{idea}. 
These samples have been created based on the $e^+ e^-$ collisions at the FCC-ee mode, featuring  $\sqrt{s} = 240$ GeV, reflecting the conditions of RUN 1.

The possible SM background processes which can mimic the DH signal are $Z/\gamma^{*}$ $\rightarrow \mu^+\mu^-$, and $\rightarrow \tau^+\tau^-$, and production of dibosons $ZZ \rightarrow 4\mu$, $ZZ \rightarrow \mu^+\mu^- + 2\nu$ and $W^{+}W^{-} \rightarrow \mu^+\mu^- + 2\nu$. 
To produce real top quark pairs, the center-of-mass energy must exceed approximately \(\sqrt{s} > 2m_t \approx 346 \, \text{GeV}\), as the top quark mass is \(m_t \approx 173 \, \text{GeV}\). Thus, at $e^+ e^-$ collider with $\sqrt{s} = 240$ GeV, the energy is too low for the process \(e^+ e^- \rightarrow t\bar{t}\) to produce on-shell top quark pairs at leading order.

These MC samples were employed to produce the SM backgrounds and to compute their corresponding cross sections times the branching ratios used in this analysis. These calculations were performed at leading order (LO), and the results are presented in Table \ref{table:tab3}.

Each of the SM background processes and signal samples was normalized according to its specific cross sections times branching ratios, and to the measured detector integrated luminosity ($\mathcal{L_{\text{int}}}$) of 10.8 ab$^{-1}$ \cite{fccee-lumi1,fccee-lumi2}.

To address potential systematic effects, we applied a uniform uncertainty of 5\% across the board, distributed as follows. In the context of circular colliders, the contribution of beamstrahlung to the uncertainty in center-of-mass energy is notably lower, at around 1\% \cite{idea}. This lower uncertainty allows for tighter constraints on the beam energy, which is particularly advantageous when determining the Higgs recoil mass.
Regarding beam energy spread (BES), it is possible to achieve measurement precision of better than 1\% by utilizing the abundant events from $e^{+}e^{-} \rightarrow \mu^{+}\mu^{-}(\gamma)$ and $e^{+}e^{-} \rightarrow e^{+}e^{-}(\gamma)$ radiative returns within the same dataset \cite{delphes-systematics}. 
For the IDEA detector concept for FCC-ee, the electromagnetic energy resolution can be effectively characterized as a function of the electron energy, ranging from 0.6\% for higher energies to 5\% as the electron energy decreases. Finally, the transverse momentum resolution is maintained at 0.3\% for a momentum of 100 GeV/c \cite{idea}.
\begin{table}
\centering
\begin {tabular} {|l|l|l|c|l|}
\hline
Process \hspace{1cm} & Deacy channel  & Generator  & {$\sigma \times \text{BR} ~(\text{fb})$} & Order \\
\hline
\hline
$Z/\gamma^{*}$ & $\mu^{+}\mu^{-}$ & Whizard & 4776.0 & LO\hspace{6cm}\\
\hline
$Z/\gamma^{*}$ & $\tau^{+}\tau^{-}$ & Whizard & 4826.0 & LO\hspace{6cm}\\
\hline
WW & $\mu^{+}\mu^{-} + 2\nu$ & Whizard & 200.6 & LO \\
\hline
ZZ & $\mu^{+}\mu^{-} + 2\nu$  & Whizard & 5.0 & LO \\
\hline
ZZ  & $4\mu$ & Whizard & 0.6 & LO \\
\hline
\end {tabular}
\caption{The SM background MC samples generated from $e^+ e^-$ collisions at the FCC-ee ($\sqrt{s} = 240$ GeV) are illustrated; process type, the event generator, the calculated cross section times branching ratios, and its order of calculation.}
\label{table:tab3}
\end{table}

The analysis revealed that the product of the LO cross section and branching ratios, calculated for various mass points of $h_{D}$ and different values of the coupling $\text{g}_{l}$, is directly proportional to $\text{g}^{2}_{l}$. Notably, the dependence on the mass of the dark Higgs is quite weak.

\section{Event selection and background reduction}
\label{section:AnSelection}

\subsection{Pre-selection of event}
\label{section:finalcuts}
This selection was employed to choose an event that contains two opposite-charge muons with low transverse momentum ($p^{\mu}_{T}$) and missing energy ($E^{miss}$) that accounts for DM particles. 
It requires both muons to meet specific preliminary criteria. \\
- $p^{\mu}_{T}$ $> 5$ GeV,\\
- $|\eta^{\mu}|$ $<$ 2.5,\\
- $\Sigma_{i} p^{i}_{T}/p^{\mu}_{T} < 0.1$. 

\begin{figure} 
\centering
\resizebox*{8.cm}{!}{\includegraphics{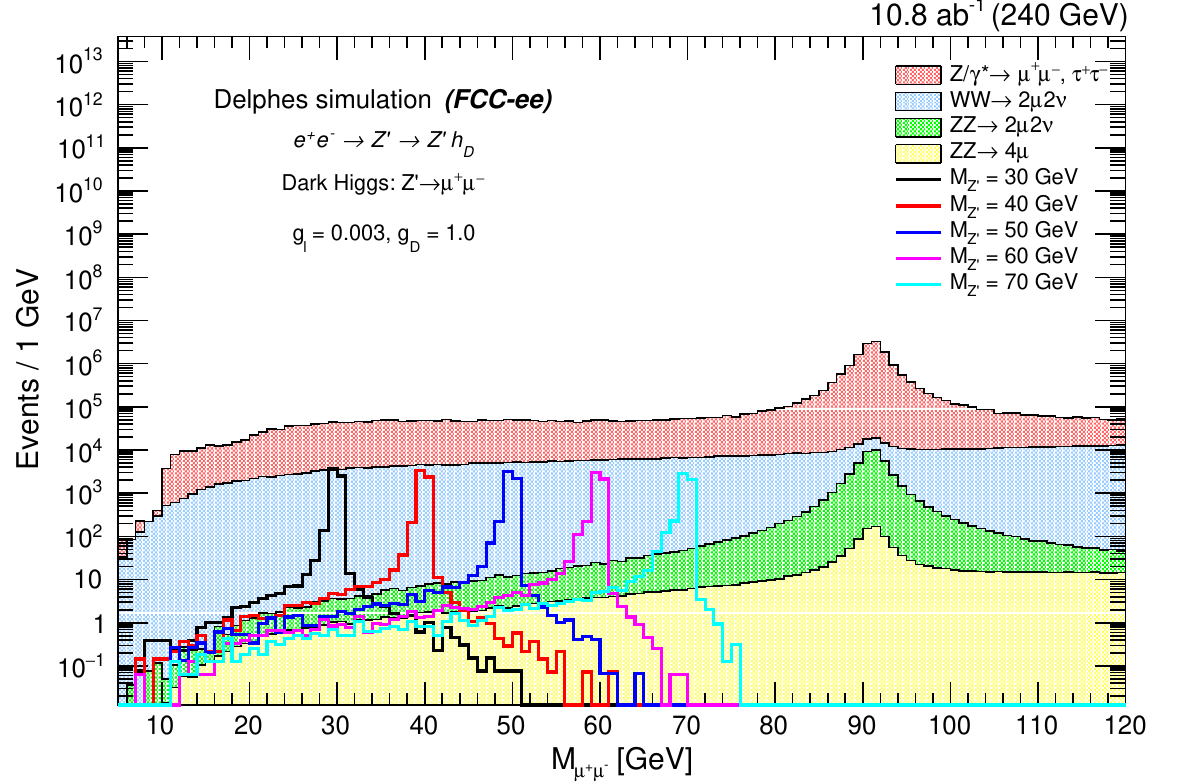}}
\caption{After applying pre-selection from Table \ref{cuts}, the dimuon invariant mass is presented for the SM background samples and various Z$^{\prime}$ mass choices from the DH scenario, with $\text{g}_{l} = 0.003$ and $\text{g}_{D} = 1.0$.}
\label{figure:fig3}
\end{figure}

In this context, $\eta^{\mu}$ denotes the pseudorapidity of the muon. 
The condition $\Sigma_{i} p^{i}_{T}/p^{\mu}_{T} < 0.1$ indicates the isolation cut, which is presented and explained in DELPHES software \cite{delphes} to reject muons from jets. 
This pre-selection criterion is illustrated in Table \ref{cuts}.

\begin{table} [h!]
    \centering
    \begin{tabular}{|c|c|}
\hline
Pre-selection & Final selection  \\
\hline
    \hline
  $p^{\mu}_{T} >$ 5 GeV &     \\
  $|\eta^{\mu}| <$ 2.5  &  Pre-selection\\
  $\Sigma_{i} p^{i}_{T}/p^{\mu}_{T} < 0.1$& \\
  $M_{\mu^{+}\mu^{-}} < 120$ GeV&\\
  \hline
  &$p_{T}^{\mu^{+}\mu^{-}} > 90.0$ GeV \\
  &$|E^{\mu^{+}\mu^{-}} - E^{\text{miss}}|/E^{\mu^{+}\mu^{-}} <$ 0.3   \\
    &$\Delta R(\mu^{+}\mu^{-}) < 1.6$\\
    &$0.9 \times M_{Z^{\prime}} < M_{\mu^{+}\mu^{-}} < M_{Z^{\prime}} + 25$\\
    \hline
    \end{tabular}
\caption{Overview of the pre-selection and final selection criteria utilized in the current study.}
    \label{cuts}
\end{table}

Figure \ref{figure:fig3} illustrates the invariant mass distribution of dimuon events meeting 
the pre-selection mentioned before. 
The SM processes are presented in colored histograms, with red showing the $Z/\gamma^{*}$ process,
cyan illustrating the $WW$ process, green for $ZZ\rightarrow 2\mu2\nu$, and yellow depicting $ZZ\rightarrow 4\mu$. 
While the signal samples from the DH scenario with Z$^{\prime}$ boson masses of 30 to 70 GeV, using fixed parameters $\text{g}_{l} = 0.003$ and $\text{g}_{D} = 1.0$, have been overlaid in different colors.

In $e^+e^-$ collisions, the total momentum and energy of final states are well understood, with only minor distortions from ISR and beam-energy spread (BES). When the decay products of the $Z^{\prime}$ boson are identified, one can calculate the momentum and energy of the dark Higgs particle, allowing us to calculate its mass using the recoil mass $(M_{rec})$ based on energy-momentum conservation, regardless of its decay mode \cite{Mrecoil}.
Then, the recoil mass is calculated from the following equation \cite{Mrecoil},
\begin{equation*}
    M_{rec} = \sqrt{s + M_{\mu^{+}\mu^{-}} - 2 \sqrt{s} E^{\mu^{+}\mu^{-}}}~.
\end{equation*}
In this equation, we used $\sqrt{s}$ for the center of mass energy, $E^{\mu^{+}\mu^{-}}$ for the energy of the dimuon pair, and $M_{\mu^{+}\mu^{-}}$ for their invariant mass. To minimize the effects of Z and SM Higgs boson ($ZH$) events, we set the recoil mass to $M_{rec} < 80$ GeV.

\begin{figure} [h!]
\centering
\resizebox*{8.cm}{!}{\includegraphics{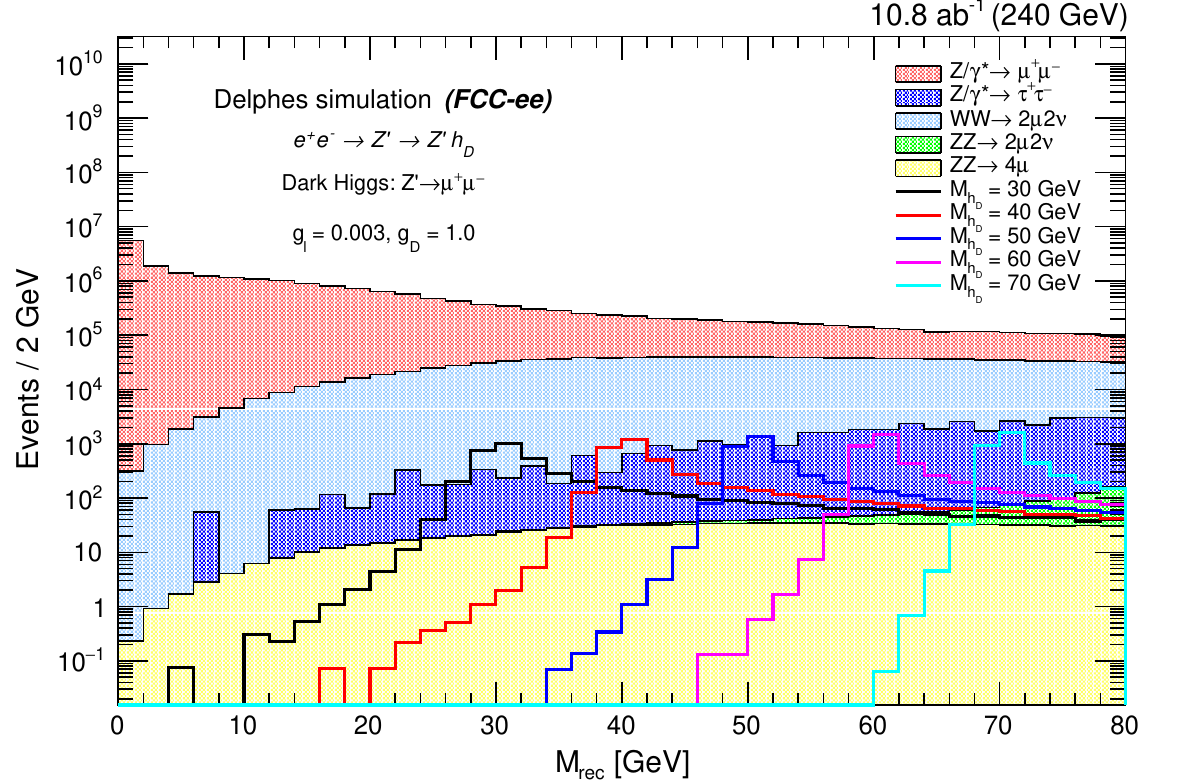}}
\caption{The recoil mass distribution for the pre-selected events detailed in Table \ref{cuts} is presented, focusing on the estimated SM backgrounds while varying the dark Higgs masses with $\text{g}_{l} = 0.003$ and $\text{g}_{D} = 1.0$.}
\label{figure:rec}
\end{figure}
In Figure \ref{figure:rec}, we illustrate the recoil mass distribution for events meeting the pre-selection criteria from Table \ref{cuts}. The histograms display estimated SM backgrounds and dark Higgs masses ($M_{h_{D}} =$ 30, 40, 50, 60, and 70 GeV), within the DH scenario, and $\text{g}_{l} = 0.003$ and $\text{g}_{D} = 1.0$.

Figures \ref{figure:fig3} and \ref{figure:rec} demonstrate that the signal samples are fully immersed in the estimated backgrounds for the entire range of the dimuon invariant mass and recoil mass. Thus, we must implement stronger criteria to separate the signal shape from the backgrounds better.
This criterion will be demonstrated and explained in the following subsection.

\subsection{Final selection of event}
\label{section:finalcuts}
We studied tighter selection cuts following four key variables. 
First, the transverse momentum of the dimuon system \( (P_{T}^{\mu^{+}\mu^{-}}) \) has to be greater than 90 GeV.
Second, the relative normalized difference between the missing energy \( (E^{\text{miss}}) \) and the energy of the dimuon system \( E^{\mu^{+}\mu^{-}}\) must be less than 0.3, defined by \( |E^{\mu^{+}\mu^{-}} - E^{\text{miss}}|/E^{\mu^{+}\mu^{-}} < 0.3 \).
Third, the angular separation \( \Delta R(\mu^{+}\mu^{-}) \) between the two muons should be less than 1.6.
Finally, since the background is broad in the dimuon mass and the detector resolution is good, imposing a cut on the dimuon invariant mass in addition to the recoil mass could greatly improve the analysis sensitivity. This could be achieved by adding the following cut on the dimuon invariant mass, thus $0.9 \times M_{Z^{\prime}} < M_{\mu^{+}\mu^{-}} < M_{Z^{\prime}} + 25$,
as recommended in \cite{R1} to suppress the non-peaking backgrounds.

\begin{figure*}
\centering
\subfigure[]{
  \includegraphics[width=75mm]{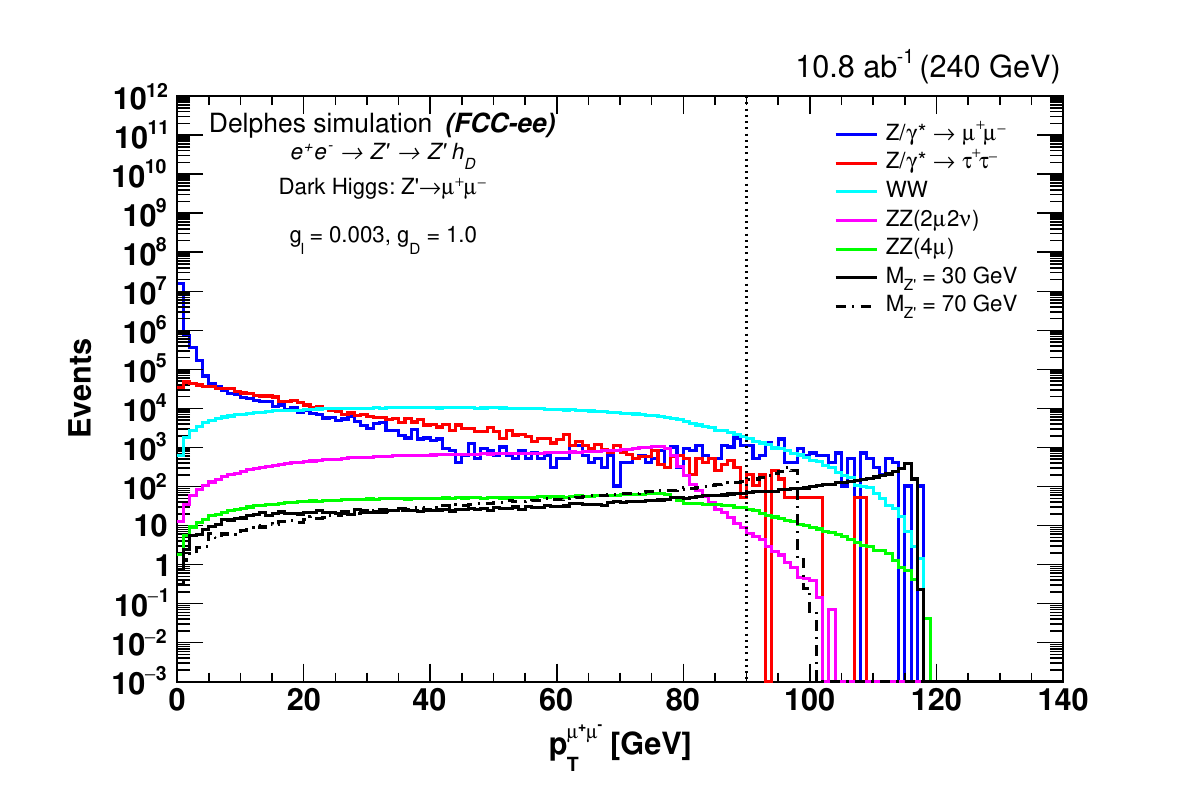}
  \label{figure:ptLeadmu}
}
\subfigure[]{
  \includegraphics[width=75mm]{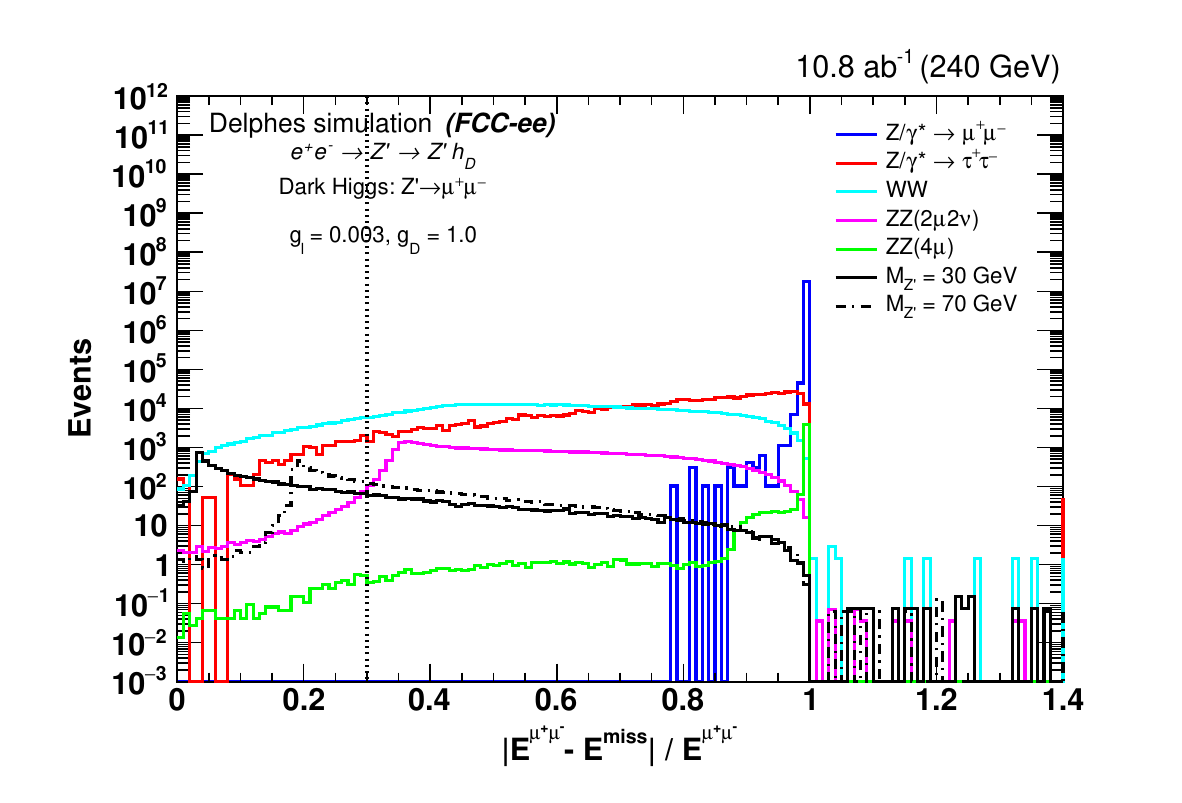}
  \label{figure:ptdiff}
}
\\
\subfigure[]{
  \includegraphics[width=75mm]{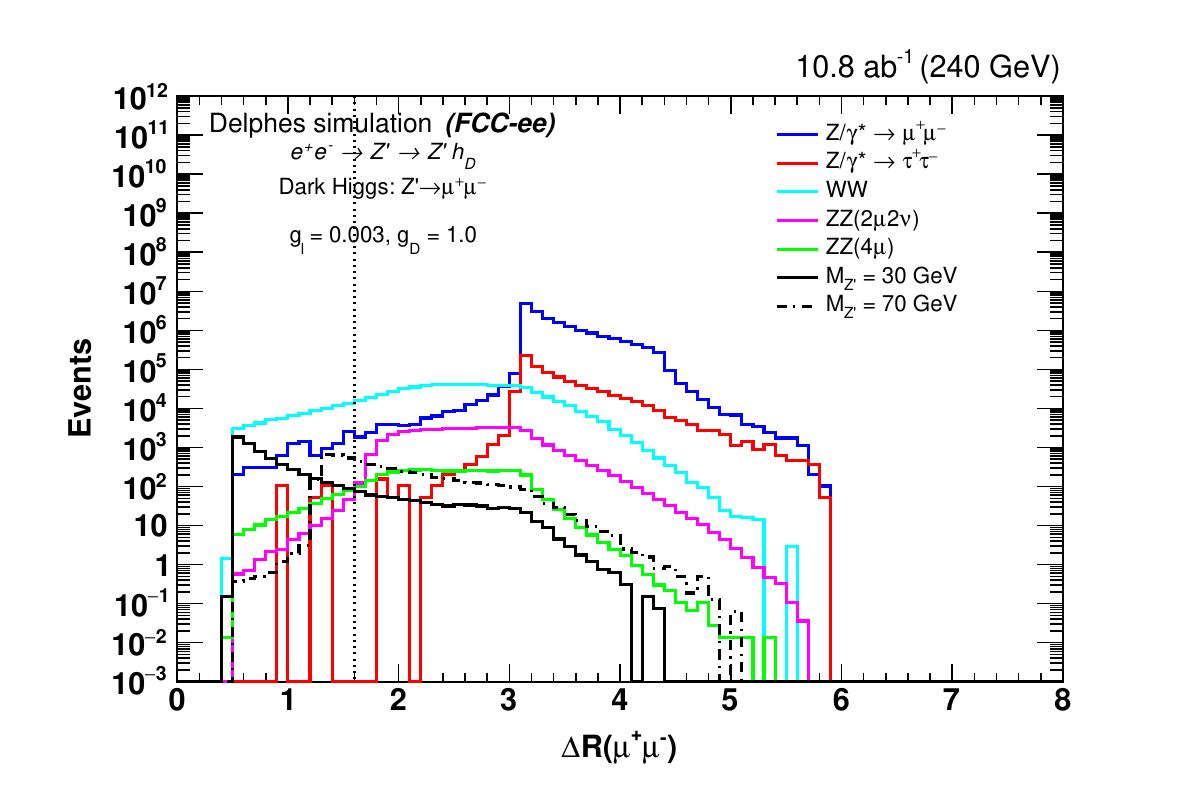}
  \label{figure:deltar}
}
\caption{The three variables distributions, for the pre-selected events mentioned in Table \ref{section:AnSelection}, include: 
$p_{T}^{\mu^{+}\mu^{-}}$ \ref{figure:ptLeadmu},
$|E^{\mu^{+}\mu^{-}} - E^{\text{miss}}|/E^{\mu^{+}\mu^{-}}$ \ref{figure:ptdiff},
$\Delta R(\mu^{+}\mu^{-})$ \ref{figure:deltar}.
These histograms represent the DH scenario for two $Z^{\prime}$ masses (30 and 70 GeV) and include SM backgrounds. The vertical black dashed lines refer to the chosen value of each cut.}
\label{figure:fig70}
\end{figure*}

For dimuon events that satisfy the pre-selection cuts, the histograms presented in Figure \ref{figure:fig70} illustrate the distributions of these four cuts for two signal choices within the DH scenario, specifically for \(M_{Z^{\prime}}\) values of 30 GeV and 70 GeV, in addition to the SM backgrounds.

The variables $p_{T}^{\mu^{+}\mu^{-}}$, 
$|E^{\mu^{+}\mu^{-}} - E^{\text{miss}}|/E^{\mu^{+}\mu^{-}}$, 
and $\Delta R(\mu^{+}\mu^{-})$ are presented in Figures \ref{figure:ptLeadmu}, \ref{figure:ptdiff}, and \ref{figure:deltar} respectively. The cut values are marked with vertical black dashed lines.
The selected cut values are determined by examining the variable distributions shown in Figure \ref{figure:fig70}.

\section{Results}
\label{section:Results}

\begin{figure*}
\centering
\subfigure[]{
  \includegraphics[width=70mm]{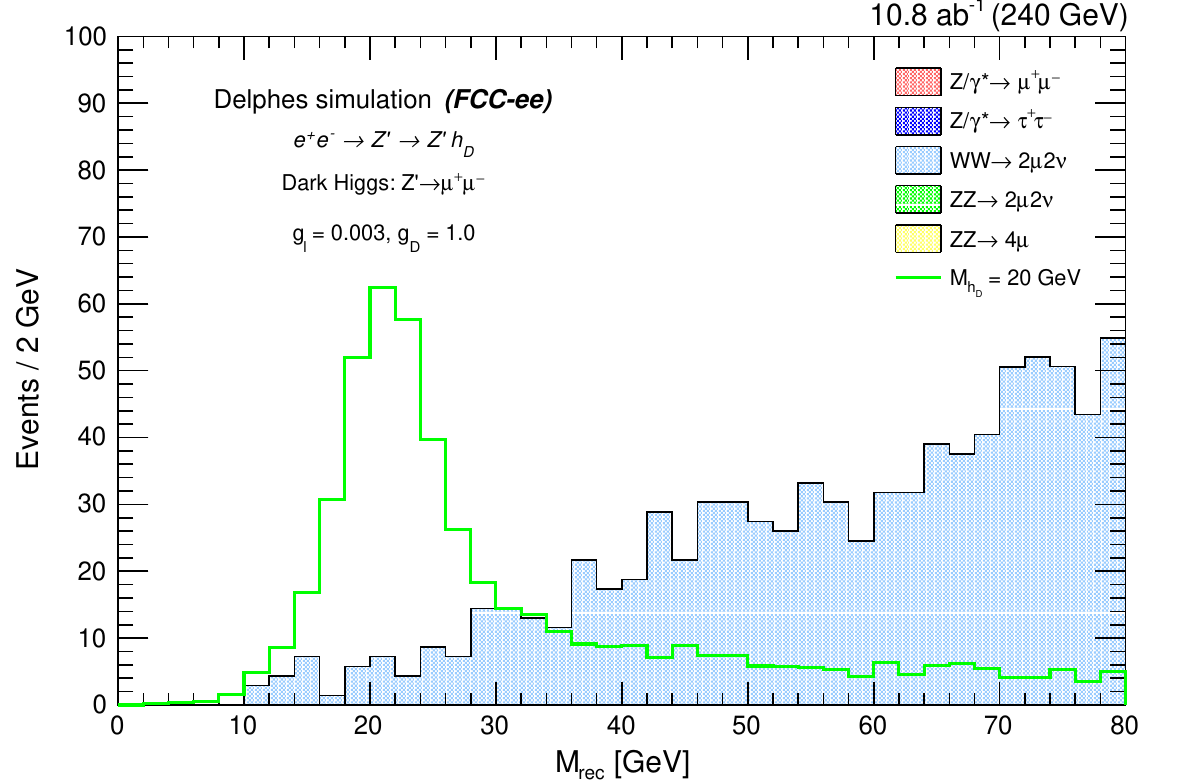}
  \label{RM20}
}
\subfigure[]{
  \includegraphics[width=70mm]{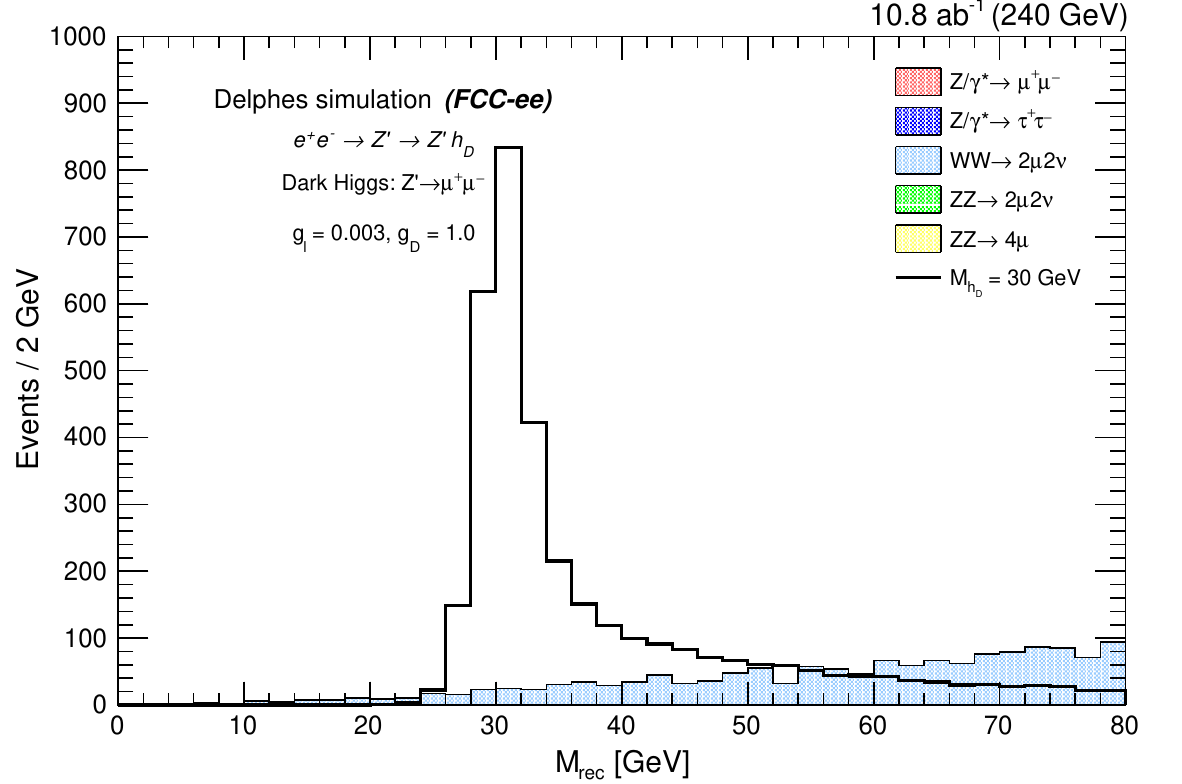}
  \label{RM30}
}
\\
  \subfigure[]{
  \includegraphics[width=70mm]{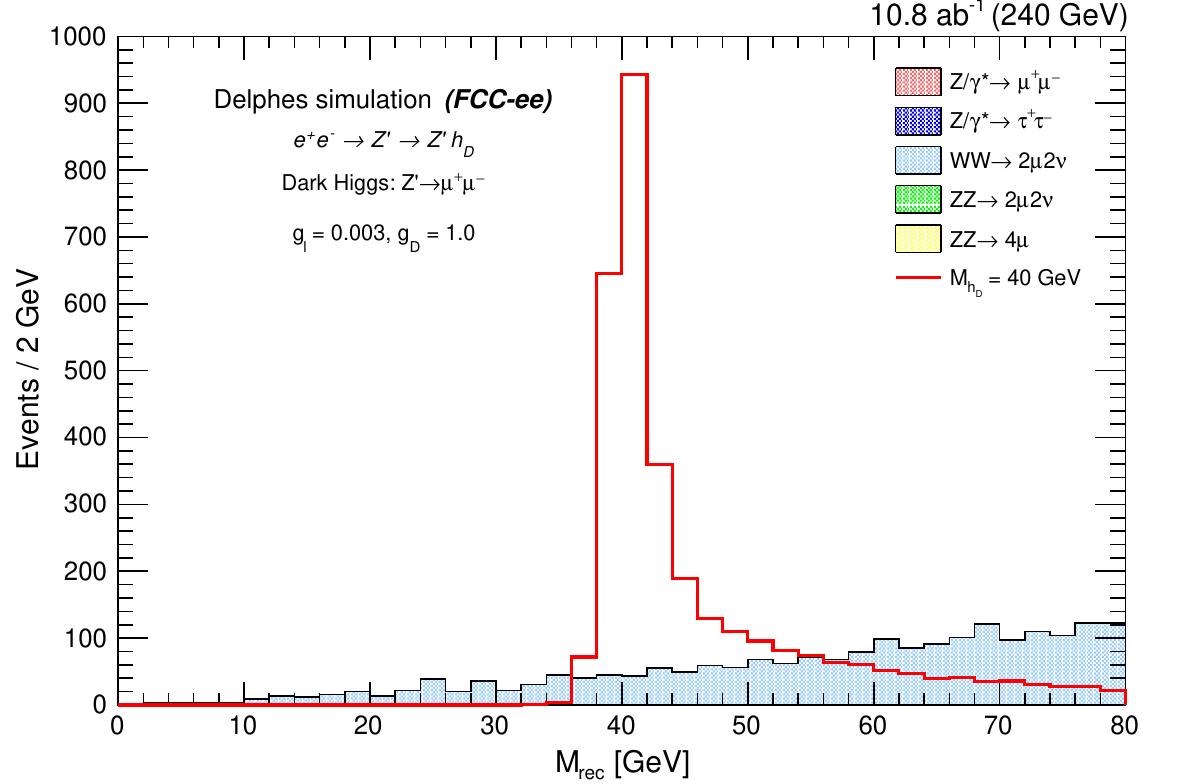}
  \label{RM40}
}
\subfigure[]{
  \includegraphics[width=70mm]{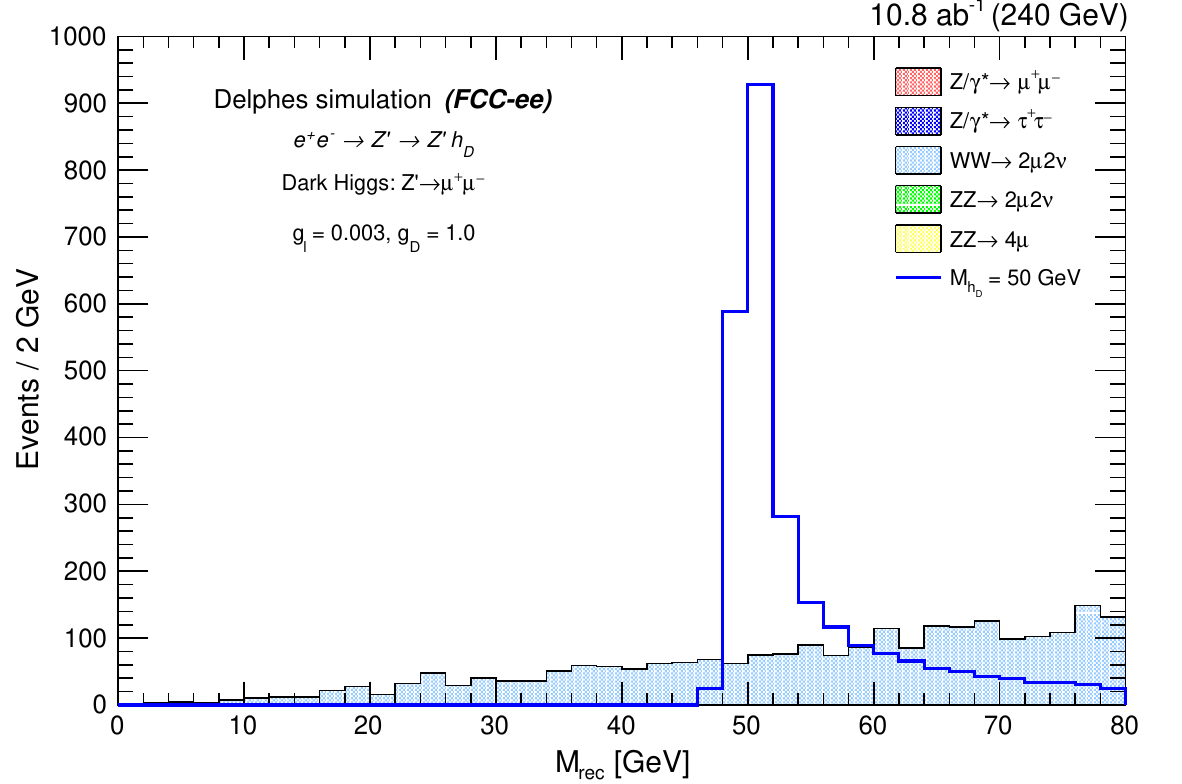}
  \label{RM50}
}

  \subfigure[]{
  \includegraphics[width=70mm]{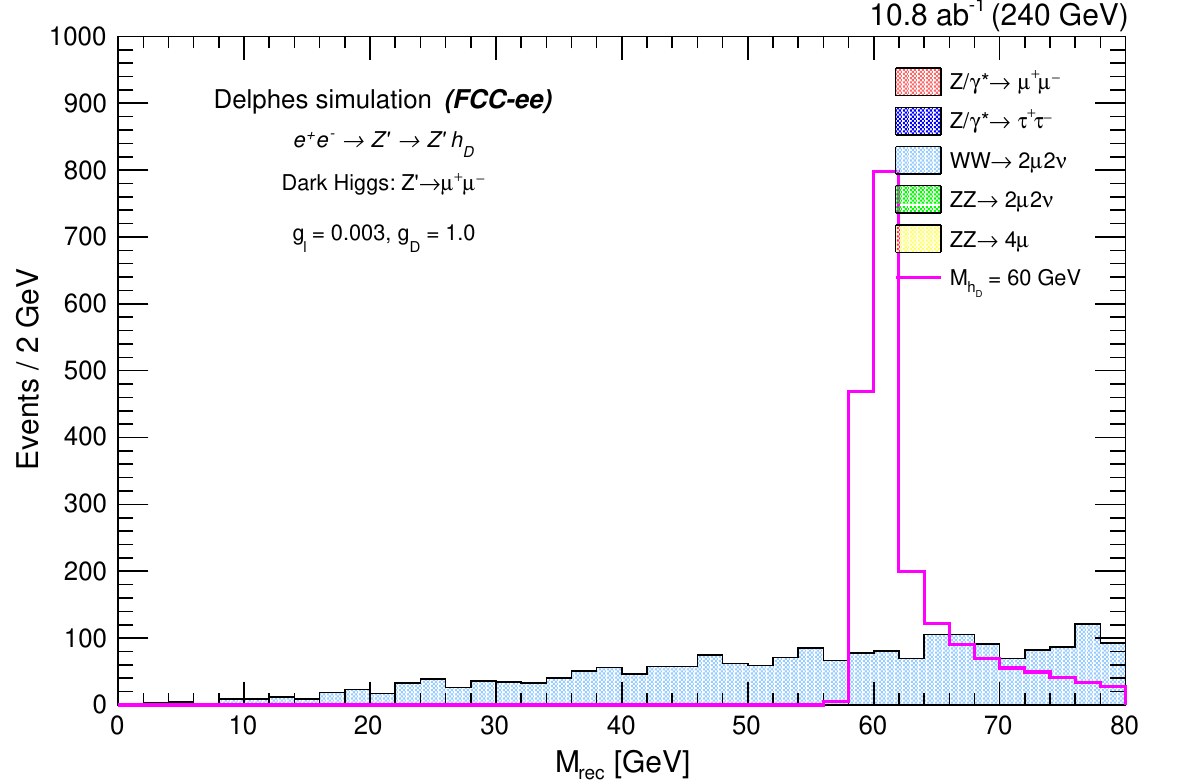}
  \label{RM60}
}
\subfigure[]{
  \includegraphics[width=70mm]{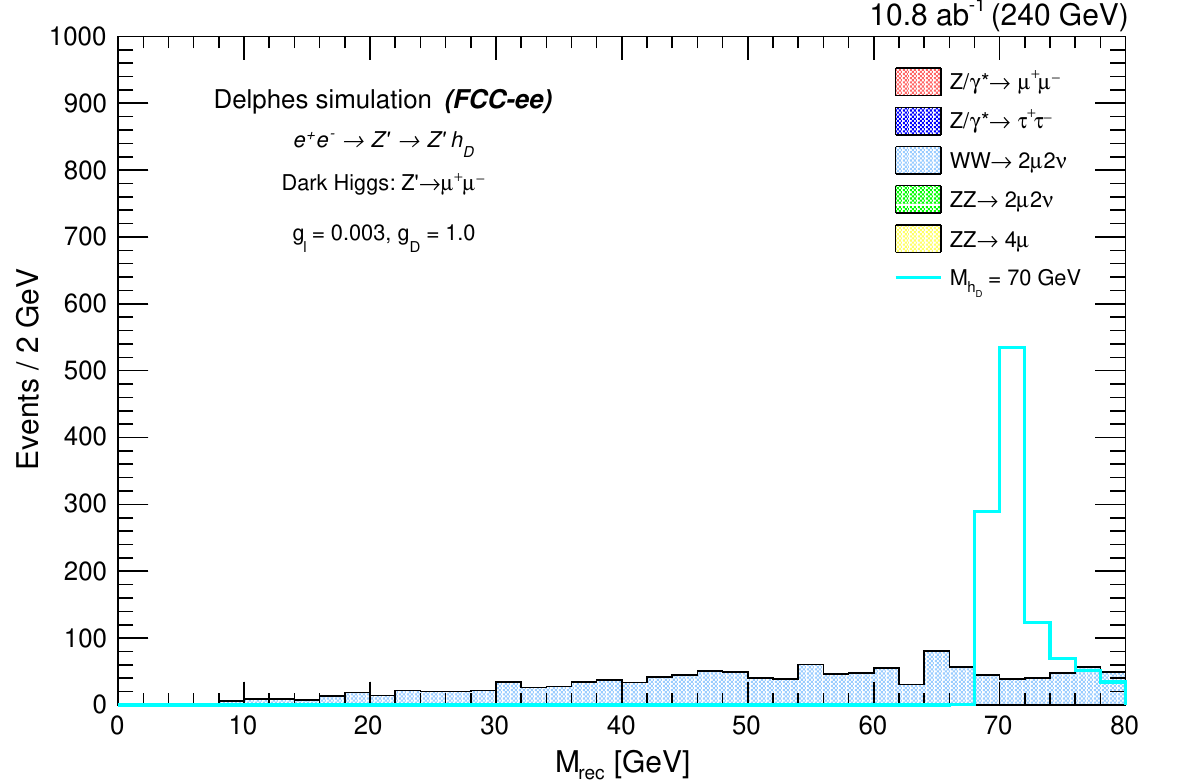}
  \label{RM70}
}
\caption{The distribution of the recoil mass for events that pass the final selection summarized in Table \ref{cuts}, the estimated SM backgrounds are presented in addition to the DH masses ($M_{h_{D}} =$ 20 \ref{RM20}, 30 \ref{RM30}, 40 \ref{RM40}, 50 \ref{RM50}, 60 \ref{RM60}, and 70 \ref{RM70} GeV), with coupling values $\text{g}_{l} = 0.003$ and $\text{g}_{D} = 1.0$.}
\label{figure:masssemifinal}
\end{figure*}


In Figure \ref{figure:masssemifinal}, we introduce the recoil mass distribution for events meeting the analysis final selection criteria mentioned in Table \ref{cuts}. The histograms display SM backgrounds and dark Higgs masses ($M_{h_{D}} =$ 30, 40, 50, 60, and 70 GeV) within the DH scenario, with parameters $\text{g}_{l} = 0.003$ and $\text{g}_{D} = 1.0$.

We utilized shape-based analysis to interpret our final results statistically, benefiting from the recoil mass distributions $(M_{rec})$ within the range $(M_{h_{D}}-5<M_{rec}<M_{h_{D}}+5)$. This range serves as an effective discriminator for events that satisfy the final selection cuts in Table \ref{cuts}.

Table \ref{table:tab8} summarizes the number of events that satisfy the final cuts across various recoil mass bins, specifically within the range $(M_{h_{D}}-5<M_{rec}<M_{h_{D}}+5)$ for the DH scenario. This is based on $\text{g}_{D} = 1.0$ and $\text{g}_{l} = 0.003$. The reported numbers incorporate all SM backgrounds, calculated for $\mathcal{L_{\text{int}}} =$ 10.8 ab$^{-1}$ at $\sqrt{s} = 240$ GeV. Additionally, the computed errors account for both statistical and systematic uncertainties.


\begin{table*}
\centering
\fontsize{5.7pt}{11pt}
\selectfont
 
\begin {tabular} {|l|c|c|c|c|c|c|}
\hline
$M_{rec}$ (GeV) & [15,25] & [25,35] & [35,45] & [45,55] & [55,65] & [65,75]  \\
\hline
\hline
$WW$ 
&$ 27.4 \pm   5.4$
&$109.8 \pm  11.8$
&$229.7 \pm  19.0$
&$355.4 \pm  25.9$
&$390.0 \pm  27.8$
&$249.9 \pm  20.2$
\\
\hline
$ZZ \rightarrow 2\mu2\nu$
&$0.0 \pm  0.0$ 
&$0.0 \pm 0.0$
&$0.0 \pm 0.0$
&$0.0 \pm 0.0$
&$0.1 \pm 0.2$
&$0.2 \pm   0.4$

\\ 
\hline
$ZZ \rightarrow 4\mu$
&$0.0 \pm 0.0$
&$0.0 \pm 0.0$
&$0.0 \pm 0.0$
&$0.1 \pm 0.2$
&$0.0 \pm 0.0$
&$0.0 \pm 0.0$
\\
\hline
Sum Bkgs 
&$ 27.4 \pm   5.4$
&$109.8 \pm  11.8$
&$229.7 \pm  19.0$ 
&$355.5 \pm  25.9$
&$390.1 \pm  27.8$
&$250.1 \pm  20.2$
\\
\hline
\hline
DH signal  
&$235.8 \pm  19.4$
&$2160.6 \pm 117.6$
&$2123.6 \pm 115.7$
&$1909.8 \pm 105.0$
&$1536.1 \pm  86.2$ 
&$985.4 \pm  58.4$
\\
\hline
\end {tabular}
\caption{The event yields that pass the final cuts are computed for various recoil mass bins $(M_{h_{D}}-5<M_{rec}<M_{h_{D}}+5)$, for all SM backgrounds 
and DH signals with $M_{h_{D}} = 20,30,...,70$ GeV for $\mathcal{L_{\text{int}}} =$ 10.8 ab$^{-1}$ at fixed values of $\text{g}_{l} = 0.003$, $\text{g}_{D} = 1.0$. 
The computed uncertainties contain statistical and systematic errors.}  
\label{table:tab8}
\end{table*}

The statistical significance ($S$) of the excess signal events in comparison to the SM background events was determined following the approximate Poisson significance formula 
\cite{R2},
\begin{equation*}
S = \sqrt{2\Big[(N_s+N_b)~\text{ln}\Big(1+\frac{N_s}{N_b}\Big) - N_s \Big]},
\end{equation*}

where $N_s$ represents the event yield for signal events and $N_b$ represents the event yield for the sum of all relevant SM background events, as taken from Table \ref{table:tab8}.

\begin{figure} [ht]
\centering
\resizebox*{8.cm}{!}{\includegraphics{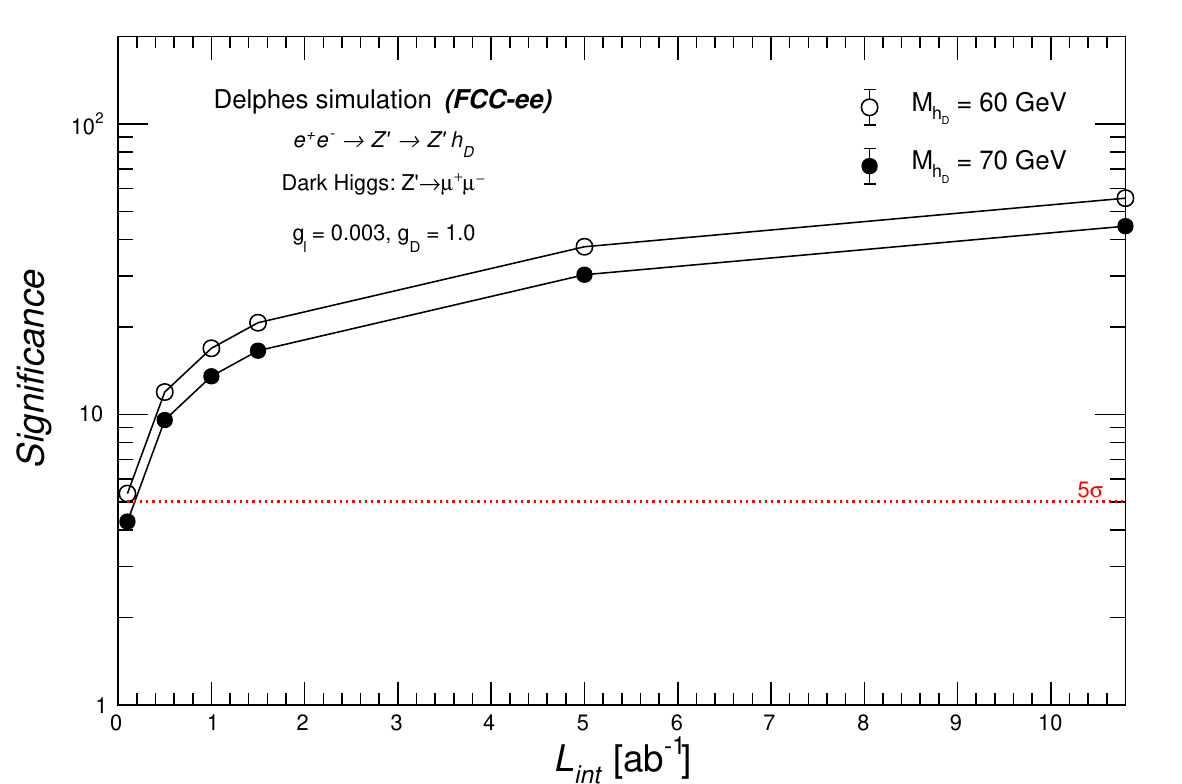}}
\caption{The calculated significances ($\it{S}$) at different values of $\mathcal{L_{\text{int}}}$, presented for $M_{h_{D}} = 60$ (open circles) and 70 (closed circles) GeV, are illustrated for events that satisfy the final cuts. 
The red dashed line represents the value of $\it{S} = 5\sigma$.}
\label{figure:significance}
\end{figure}

Figure \ref{figure:significance} shows significance $(S)$ versus the FCC-ee's integrated luminosity ($\mathcal{L_{\text{int}}}$) computed for dark Higgs masses by taking \(M_{h_{D}} = 60\) and \(70\) GeV, based on events pass the final cuts. 
The dashed red line refers to \(S = 5\sigma\). 
For $\mathcal{L_{\text{int}}} > 100$ fb\(^{-1}\) we can exceed 5$\sigma$ discovery for $M_{h_{D}} =$ 60, and 70 GeV.

We statistically analyzed our results using the profile likelihood method and used a statistical test based on the $M_{rec}$ histograms. Employing the modified frequentist construction CLs \cite{R58, R59} and the asymptotic approximation \cite{R2}. Thus, we established exclusion limits on the signal cross sections times the branching fraction Br($Z^{\prime}$ $\rightarrow \mu\mu$) at a 95\% CL, treating the systematic uncertainties as parameters of nuisance.

\begin{figure}
\centering
  \resizebox*{8.0cm}{!}{\includegraphics{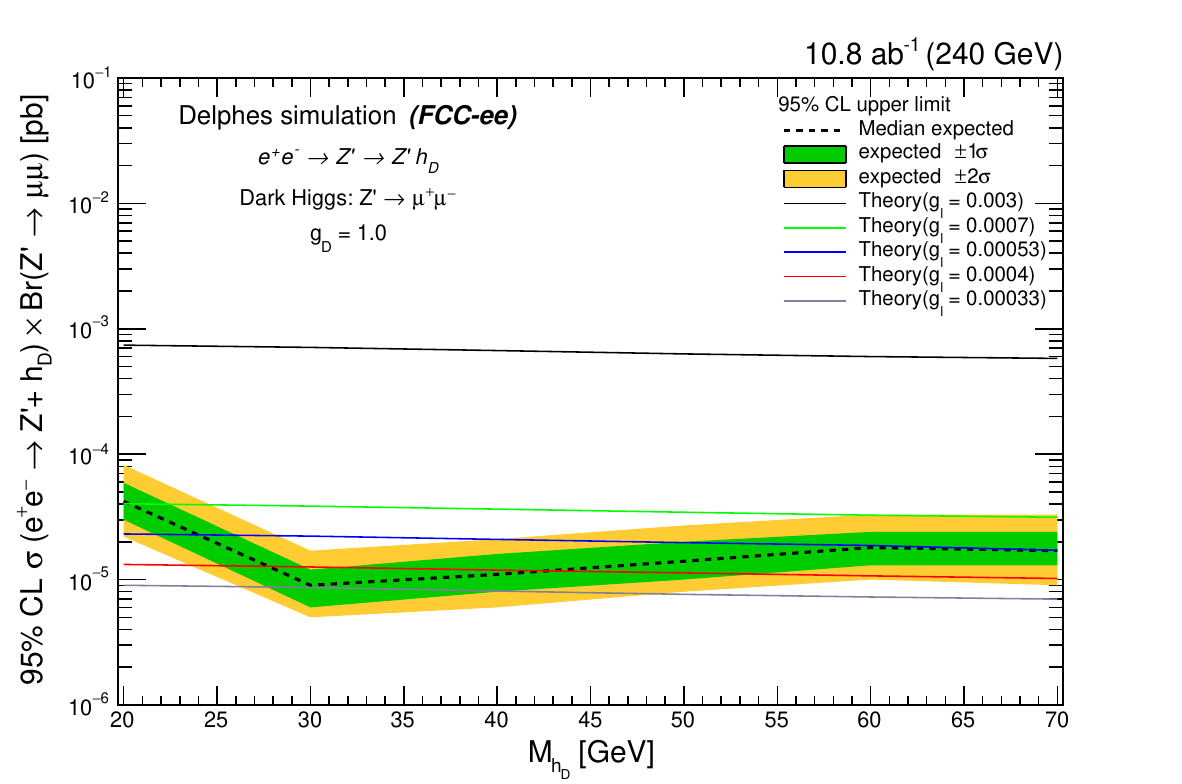}}
\caption{The expected limits, at 95\% CL, on the cross section times the branching ratio versus $M_{h_D}$ with muonic Z$^{\prime}$ decay. The solid colored lines represent the DH scenario for several choices of the coupling $\text{g}_{l}$.}
\label{figure:fig7}
\end{figure}

Figure \ref{figure:fig7} shows the expected 95\% CL upper limit on the cross section times Br($Z^{\prime}$ $\rightarrow \mu\mu$) for the DH scenario. This result is demonstrated for \(\text{g}_{D} = 1.0\), and derived from an $\mathcal{L_{\text{int}}}$ of 10.8 ab$^{-1}$ at \(\sqrt{s} = 240\) GeV. These limits are represented by colored solid lines for various coupling constants \(\text{g}_{l}\): 0.00033, 0.0004, 0.00053, 0.0007, and 0.003 GeV.

\begin{figure}
\centering
  \resizebox*{8.cm}{!}{\includegraphics{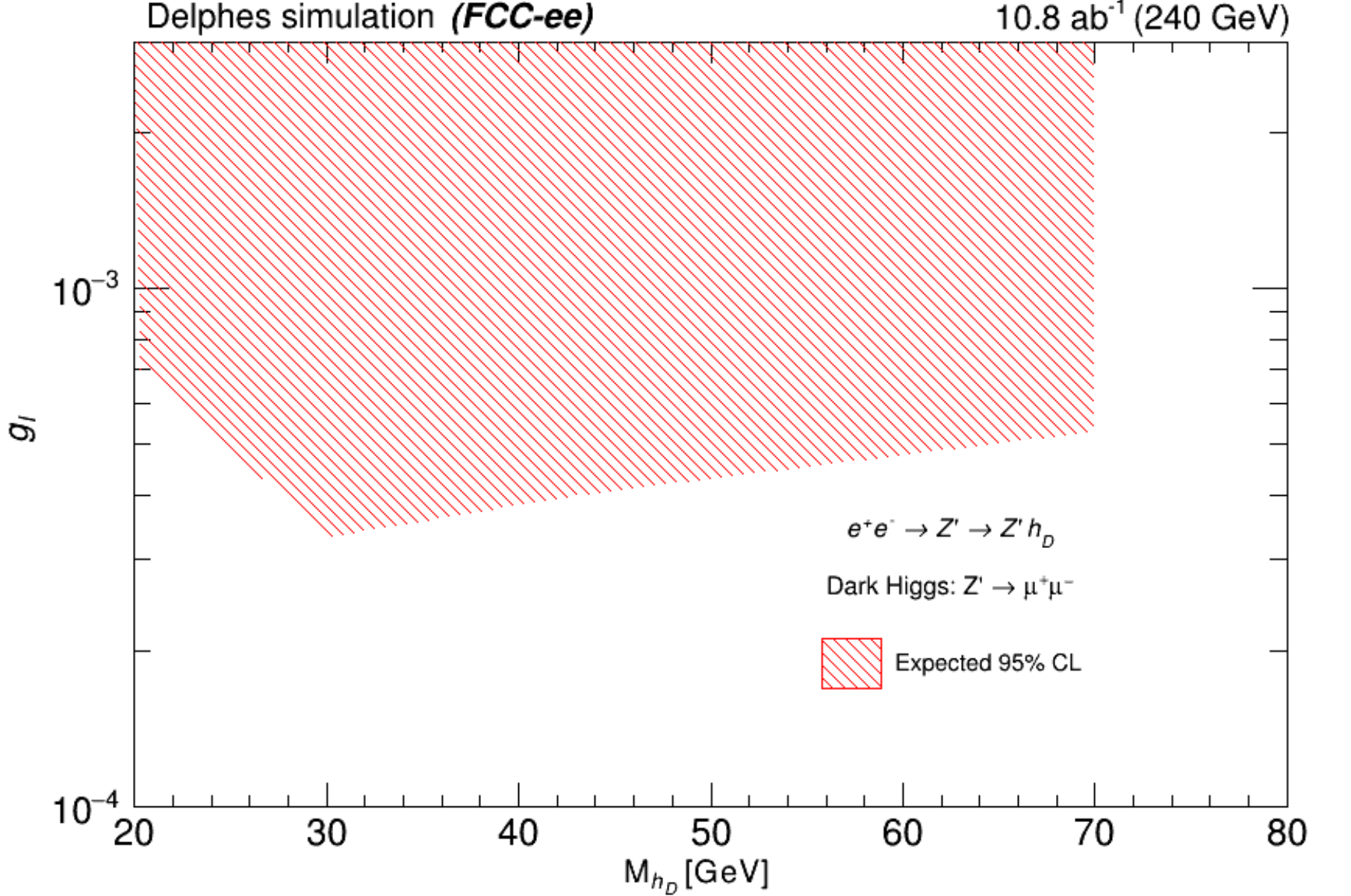}}
    \caption{The excluded parameter space region in the DH scenario parameters ($M_{h_{D}}$ and $\texttt{g}_{l}$) is shown.  
    The red region is excluded at 95\% CL for the coupling constant $\text{g}_{D} = 1.0$ at $\sqrt{s} = 240$ GeV and an $\mathcal{L_{\text{int}}} =$ 10.8 ab\(^{-1}\).}
  \label{figure:fig8}
\end{figure}

Figure \ref{figure:fig8} depicts the excluded parameter space at the 95\% CL in the DH scenario for \(\text{g}_{D} = 1.0\), $\mathcal{L_{\text{int}}} =$ 10.8 ab\(^{-1}\), and \(\sqrt{s} = 240\) GeV.
The maximum excluded \(\text{g}_{l}\) is 0.003 for \(M_{h_{D}}\) values between 20 and 70 GeV, while the minimum excluded \(\text{g}_{l}\) is 0.00033 for \(M_{h_{D}} = 30\) GeV.

\section{Summary}
\label{section:Summary}
The Foreseen circular collider in $e^+ e^-$ collision mode (FCC-ee), which will take place at CERN, 
will pave the way to observe new particles beyond the Standard Model.
This provides a way to identify undiscovered particles such as dark matter, a dark Higgs, and extra neutral gauge bosons.

Our research analyzed the recoil mass distribution, which is characterized by low mass through the Mono-Z$^{\prime}$ portal (i.e., $M_{rec} < 80$ GeV) using simulated Monte Carlo data from FCC-ee. The samples were generated based on $e^+ e^-$ collisions with $\sqrt{s} = 240$ GeV, including both signal and SM background events, aligned with expectations for FCC-ee Run 1 with $\mathcal{L_{\text{int}}} =$ 10.8 ab\(^{-1}\). 
We focused on the possible generation of a dark Higgs boson in addition to a \( Z^{\prime} \) boson at the FCC-ee. Specifically, we examined the muonic decay mode of the \( Z^{\prime} \), taking into account the coupling values \( \text{g}_{l} \leq 0.003 \) and \( \text{g}_{D} = 1.0 \).

In this analysis, we proposed powerful cuts that suppressed all SM backgrounds relevant to this study except WW, which we strongly reduced, while maintaining signal strength. This enabled us to distinguish signal from SM background events more effectively. Using the analysis final cuts, we found that dark Higgs signals with a mass \(M_{h_{D}} > 20\) GeV can much exceed the 5$\sigma$ discovery threshold for $\mathcal{L_{\text{int}}} =$ 10.8 ab\(^{-1}\) when the coupling constants are fixed to \(\text{g}_{D} = 1.0\) and \(\text{g}_{l} = 0.003\).

Finally, in case the dark Higgs boson is not seen at the FCC-ee, we set upper limits on its mass ($M_{h_{D}}$) at 95\% CL for the muonic decay of the on-shell \( Z' \), with the use of various choices of $\text{g}_{l}$ with $\text{g}_{D} = 1.0$. 
The mass of the $h_{D}$ particle, between 20 and 70 GeV, can be excluded for a coupling strength of $\text{g}_{l} = 0.003$ with $\mathcal{L_{\text{int}}} =$ 10.8 ab$^{-1}$ at $\sqrt{s} = 240$ GeV. 
For $\text{g}_{l} < 0.00033$, the FCC-ee will lose its sensitivity to the DH scenario in the framework 
of the Mono-Z$^{\prime}$ portal characterized by these precise parameters.

Previous \(e^+ e^-\) colliders, including the LEP working group, updated the limits for the mass of a Higgs boson that decays invisibly using data from approximately 189-209 pb\(^{-1}\) \cite{LEP-Higgs}. 
No evidence was found for the invisible decay of Higgs bosons originating from the decay of a Z boson to hadrons, electrons, or muons (i.e., $hZ \rightarrow \text{inv.} + q\bar{q} / e^{+}e^{-} / \mu^{+}\mu^{-}$). Thus, the mass range between 50 and 112.1 GeV from the combination of hadronic and leptonic analyses has been excluded at 95\% CL. Our analysis lowers the limit to 20 GeV.

Additionally, the current study explores a specific dark Higgs scenario and provides a comprehensive phenomenological analysis. The sensitivities obtained in this study are expected to have broader implications for models that introduce new gauge-boson mediators linking the SM to a hidden sector. 
Since mediator's production rate and decay characteristics, our findings can be interpreted in simplified models or effective field theory (EFT) frameworks by rescaling the theory (EFT) frameworks by rescaling couplings and branching ratios accordingly.  
Notably, our analysis within the mono-\(Z'\) framework indicates that the sensitivity of the FCC-ee disappears for \( g_l \lesssim 3.3 \times 10^{-4} \). 
This result aligns with FCC-ee studies on leptophilic \( Z' \) bosons \cite{Zprime-couplin-to-ll}, which achieved sensitivities of \( 10^{-4} \) to \( 10^{-3} \) based on the mediator's mass and decay topology.

The findings on the electron-philic decay of on-shell \( Z' \) would align with this study's results, assuming that the detector resolution and reconstruction efficiencies for electrons and muons are similar.


\begin{acknowledgments}
The author of this paper would like to thank Tongyan Lin, co-author of \cite{R1}, for providing us with the UFO model files, helping us to generate the signal events, and cross-checking the results.
In addition, this paper is based on work supported by the Science, Technology, and Innovation Funding Authority (STDF) under grant number 48289. 
\end{acknowledgments}


\end{document}